# Towards Live 3D Reconstruction from Wearable Video: An Evaluation of V-SLAM, NeRF, and Videogrammetry Techniques

**David Ramirez, Suren Jayasuriya, Andreas Spanias**
**SenSIP Center, Arizona State University, Fulton Schools of Engineering**
**Tempe, Arizona**
**dframire@asu.edu, sjayasur@asu.edu, spanias@asu.edu**

## ABSTRACT

Mixed reality (MR) is a key technology which promises to change the future of warfare. An MR hybrid of physical outdoor environments and virtual military training will enable engagements with long distance enemies, both real and simulated. To enable this technology, a large-scale 3D model of a physical environment must be maintained based on live sensor observations. 3D reconstruction algorithms should utilize the low cost and pervasiveness of video camera sensors, from both overhead and soldier-level perspectives. Mapping speed and 3D quality can be balanced to enable live MR training in dynamic environments. Given these requirements, we survey several 3D reconstruction algorithms for large-scale mapping for military applications given only live video. We measure 3D reconstruction performance from common structure from motion, visual-SLAM, and photogrammetry techniques. This includes the open source algorithms COLMAP, ORB-SLAM3, and NeRF using Instant-NGP. We utilize the autonomous driving academic benchmark KITTI, which includes both dashboard camera video and lidar produced 3D ground truth. With the KITTI data, our primary contribution is a quantitative evaluation of 3D reconstruction computational speed when considering live video.

## ABOUT THE AUTHORS

**David Ramirez** is a Ph.D. Candidate in Computer Engineering at Arizona State University (ASU) and a Senior Staff Data Scientist at Maxar. He received Master’s and Bachelor’s degrees in Electrical Engineering, with a focus in digital signal processing. He has applied deep learning since 2017, contributing to many military research interests. David previously served in the United States Marine Corps, supporting peaceful military operations in 11 countries. His research interests focus on computer vision, deep learning, mixed-reality, and military training.

**Suren Jayasuriya Ph.D.** is an Assistant Professor at ASU, in the School of Arts, Media and Engineering (AME) and Electrical, Computer and Energy Engineering (ECEE). Before this, he was a postdoctoral fellow at the Robotics Institute at Carnegie Mellon University. He received his doctorate in 2017 from the ECE Department at Cornell University, and a bachelor's in mathematics and in philosophy from the University of Pittsburgh in 2012. His research focuses on designing new types of computational cameras, systems, and visual computing algorithms that can extract and understand more information from the world around us.

**Andreas Spanias Ph.D.** is Professor in the School of Electrical, Computer, and Energy Engineering at ASU. He is the founder and director of the Sensor Signal and Information Processing (SenSIP) center and industry consortium, a National Science Foundation (NSF) Industry/University Cooperative Research Center (I/UCRC). His research interests are in the areas of adaptive signal processing, speech processing, machine learning and sensor systems. He is a Fellow of the IEEE and a Senior Member of the National Academy of Inventors.

# Towards Live 3D Reconstruction from Wearable Video: An Evaluation of V-SLAM, NeRF, and Videogrammetry Techniques

**David Ramirez, Suren Jayasuriya, Andreas Spanias**
**SenSIP Center, Arizona State University, Fulton Schools of Engineering**
**Tempe, Arizona**
**dframire@asu.edu, sjayasur@asu.edu, spanias@asu.edu**

## INTRODUCTION

Mixed reality (MR), combining a physical view of the world and interaction with virtual objects, is a key technology which promises to change the future of warfare and military training. The ability to overlay relevant and timely information into a heads-up display will enable new opportunities to train soldiers through live outdoor tactical training exercises. An augmented training battlefield will allow long-distance engagements with both real and simulated opposing forces. For live military training over a large area, it becomes necessary to track not only the active agents but also the evolving physical battlespace itself. For this reason, a 3D model of the world must be maintained based on live sensor observations. Using augmented reality (AR) headsets with visual overlay, training-weapon systems, and video sensors, a digital twin model of the world can be mapped and updated in real-time. The physical battlespace and mirrored digital twin are synchronized, as shown in Figure 1. Force-on-force training engagements can be supplemented with simulated agents and remote-controlled actors added via the AR visual overlay. When a “digital bullet” is fired, the trajectory must be affected by physical intervening cover. For increased immersiveness of the training, a virtual bullet may impact near a soldier, releasing dust, sparks, shrapnel, and other virtual effects visible through the AR viewscreen. The digital twin battlefield is both a tool for tracking human combatants in the dynamic landscape and a map of the world for real-time and after-action analysis. Video of the physical environment and the digital twin model are recorded allowing for review. All these high-value objectives require a digital twin of the battlefield, captured through live video. We explore algorithms for 3D reconstruction to enable this goal.

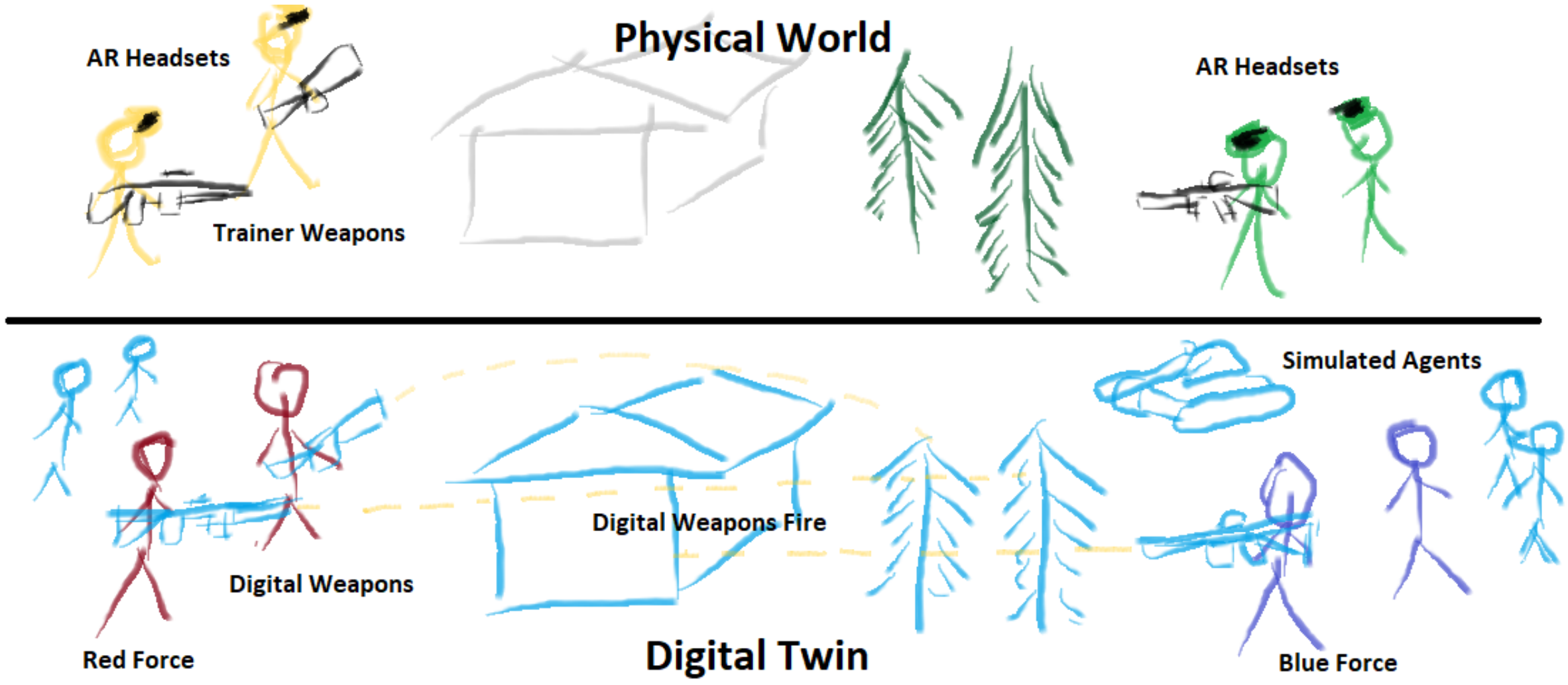

**Figure 1. MR military training will include physical environment, human soldiers, and simulated effects.**

Within this study, we define and differentiate several overlapping technologies:

- **Augmented Reality (AR)**: Human perspective superimposed with virtual objects.
- **Digital Twin**: 3D model of the physical world updated from real-time sources.
- **Mixed Reality (MR)**: Digital objects synchronized with twins are displayed via AR.

The three technologies defined above feed into each other to enhance training immersion and situational awareness. Of these concepts, AR is the most simple to implement. Virtual objects can be placed into view of a soldier without a complete model of the world. Enabling multiple human agents to view the same virtual object requires a consistent map across all views and a substantial leap in complexity. Despite the challenge, tracking a scale 3D digital twin of the world is nearly within reach for civilian applications of indoor and city environments. Real-time techniques for flat-surface detection, room mapping, and built-structure modeling are quickly advancing (Patel et al., 2020). Augmenting the digital twin world with virtual objects and projecting these back into view is possible with AR headsets (Tu et al., 2007). Our team explores the remaining challenges and potential solutions for large-scale outdoor environments.

MR relies upon several high-level building blocks. We group these into several successive steps:

1. **Agent in the Physical World**: A human or robotic agent is moving in the physical world.
2. **View of the World**: A sensor is capturing the agent's view (we assume a video camera).
3. **Location of Agent View**: Precision physical position must be measured and calculated.
4. **3D into Digital Twin:** A model of the world is constructed from sensor measurements.
5. **Large-Scale Digital Twin:** Many views contribute to the continuously updating model.
6. **Virtual Objects in World:** Objects are inserted to impact the agent's decision making.
7. **Effect on Physical Agent:** An impact on the agent's senses occurs, (we explore visual).

**Challenges in Practice**

If the above steps can be successfully implemented, an exciting new potential for MR for military training will be enabled. Knowing the true location of a sensor's point of view is the first hurdle. Due to measurement errors in GPS and inertial measurement units (IMU), visual odometry techniques for localization are necessary for improved placement of a user into the battlefield's digital twin. As for 3D mapping, even though small form factor time-of-flight sensors are now available on the latest iPhone™, structured light sensors and lidar are often too limited by range and power. To proliferate MR technology, 3D reconstruction must utilize the low cost and widespread use of video camera sensors (Shanthamallu et al, 2017), both flying overhead and soldier-level perspectives. Wearable video recording devices may include helmet-wearable and weapon-mounted cameras. Fortunately, consumer off-the-shelf action cameras offer a very capable live sensor, even including wireless video streaming to networked devices. Although the cost of hardware is now ripe for mass adoption, computer vision algorithms are still maturing. A limiting factor today is the speed of photography-based mapping technologies, known as photogrammetry or videogrammetry. Offline methods can produce high-quality 3D models within minutes, but not in real-time. Mapping speed and 3D quality must be balanced to enable live hybrid training in dynamic environments. Besides the capture of 3D and construction of the digital twin, sharing this world between AR users is another challenge. In civilian applications with 5G capable networks, video streaming is no longer a challenge. This is not the case for simulated battlefield environments at the network edge (Braun et al., 2019; Braun et al., 2020; Iqbal et al., 2020). With the above solved, it is still necessary to resolve differences between different observations. The natural environment includes weather effects, swaying trees and grass, and constantly varying lighting conditions. These problems in practice are summarized below:

- **Measuring View Location**: GPS and IMU have limitations requiring visual odometry.
- **3D Capture of Physical World**: Lidar and photogrammetry both have limitations.
- **Computational Tradeoff:** Performance is limited by electronics size, weight, and power.
- **Wireless Network**: Transmission of 3D data at scale is unsolved at the battlefield edge.
- **Sharing Between Agents**: Resolving differences between observations is challenging.

3D mapping of the outdoor world is still a substantial unsolved challenge required to enable MR at scale. Given these requirements, we review several openly available 3D reconstruction pipelines for mapping given only live video, including neural radiance fields (Mildenhall et al., 2020), ORB-SLAM3 (Campos et al., 2021), and COLMAP (Schonberger & Frahm, 2016). In this survey, we review the practical uses of photogrammetry and videogrammetry, relevant military AR applications, relevant technical foundations, and several algorithms. Finally we present our findings of 3D reconstruction algorithm performance regarding both qualitative accuracy and speed. These topics support a complete understanding of the emerging technologies and their limitations in practice.

## CONTEXT

Civilian companies and open source research have made substantial advances in 3D reconstruction algorithms within the last decade. This makes capturing and digitizing the world much simpler. An accurate and rapidly updated 3D model of the world is an important aspect for large-scale MR wargames. Few challenges remain before mass adoption is possible for civilian businesses and military use cases.

### Civilian Applications

Civilian technology companies are utilizing 3D reconstruction for consumer applications and entertainment use cases: both the photogrammetry technique for 3D capture as well as creation of a digital twin model.

These related technologies have found increasing use in 3D mapping of the world at scale. Consumer drones increasingly use photogrammetry, instead of measuring the 3D world with lidar. This competing laser-ranging technology has limitations including sensor range, measurement accuracy, cost, size, weight, and power. A quadcopter drone can take high-resolution photographs at regular intervals to survey an environment, with or without the addition of lidar. This practice is increasingly common in various industries, including surveying, forestry, geology, architecture, real estate, open-pit mining, and construction. For large-scale applications, satellites and high altitude drones are instead the collection platform of convenience. These lack the fine 3D detail necessary to build a realistic digital twin model. Furthermore, flying platforms fail to capture the environment when there is intervening cover, such as below roof overhangs, under thick tree canopies, and below cloud cover. Companies are investing to capture metropolitan cities at massive scale. A new generation of AR smartphone applications is near for navigation, games, retail, travel, and more.

In years past, 3D reconstruction was only accessible by researchers and industry specialists. Recently, Epic Games, the company that produces the Unreal Engine for digital graphics rendering, has removed barriers to access this technology. They recently purchased the Capturing Reality software company, best known for the industry-leading RealityCapture software for photogrammetry. This combined entity, as well as other competing technology companies, now offer smartphone applications for capturing the world at small scale. This is useful for 3D modeling individual objects and structures. Due to this new ease of use, the online shopping and real-estate industries may soon include 3D views of goods and properties for sale. Consumers and professionals alike are capturing aspects of the real world large and small and offering these as assets for digital worlds. These 3D assets are rarely realized in real-time, which is a requirement for a relevant digital twin 3D map of the world as we define it.

### Military Applications

AR at scale is not easy, or else it would have been solved by now. The uncertain state of AR for U.S. ground military applications is characterized by the Integrated Visual Augmentation System (IVAS) development effort (TAK, 2020). IVAS is designed from the Microsoft Hololens™ 2 AR system. It attempts to integrate many sensors, including one thermal, two low-light/night-vision, one RGB color, and four stereo cameras for head-tracking and depth sensing. The field of view offered by this AR system is unlike anything else currently available: wide enough to add relevant AR visuals even at the periphery of human vision. This comes at the cost of power usage and battery life, which do not currently meet contract requirements. IVAS, when ready, is likely to enhance situational awareness, land navigation, and coordination for coalition ground forces, but a constantly updating digital twin map of the world is less likely. It remains to be seen if IVAS will reach the lofty goals for MR military training for force-on-force firefights. Improvements in technology are necessary before MR can be enabled at scale.

Even with these setbacks, AR, digital twins, and MR remain valuable goals for the future of warfare. Given only sensors and no AR visual overlay, there are still substantial benefits for maintaining a 3D digital twin model of the battlefield which tracks and updates from live video. The benefits from 3D reconstruction include:

- **Location**: Determine the position of each video sensor relative to the digital twin map.
- **Bearing**: Determine the angle direction of the point of view relative to the world.
- **3D Map**: Track and update imprecise level of detail of terrain and sources of cover.
- **3D Render**: Display a high-definition reconstruction of the battlefield.
- **Object Tracking**: Track movement of non-stationary dynamic world elements.

The recent I/ITSEC publications below detail the remaining challenges for 3D reconstruction and MR for military training applications. Chen et al. (2021) demonstrated the process of creating a 3D digital model of a large outdoor environment. Using photogrammetry and LIDAR from overhead drones, they modeled an outdoor landscape and the built environment. This offline process leveraged One World Terrain (OWT) as a dataset for comparison. When detail was unavailable from OWT, the Unreal Engine digital twin model filled the world with procedurally generated natural terrain. This offline process created a digital world suitable for virtual reality. In contrast, the Affordable Augmented Reality Fire Support Team Trainer, as described by Sullivan et al. (2021), was focused on live, outdoor, and long range AR training against simulated opponents. This system demonstrated the positional navigation accuracy of both differentiable-GPS positions and stereo visual inertial odometry provided by consumer off the shelf hardware. In other relevant work, it was shown that dynamic objects in the physical world, including human beings, must be tracked for a functioning AR system. Martin et al. (2020) demonstrated the importance of a digital twin model of the world which occluded AR objects placed into and behind physical objects in the environment. Moralez et al. (2021) described detecting objects from the physical world and replicating these as avatars in the digital twin. Localization, 3D mapping, 3D modeling, and object tracking can all be accomplished with these I/ITSEC published techniques. With the advent of these and other algorithms, few software challenges remain to be solved to enable a large-scale AR military training system. Perhaps the only remaining challenges are speed and consistency, which we evaluate in the following sections.

## METHODS

Photogrammetry, videogrammetry, 3D reconstruction, structure from motion, visual simultaneous localization and mapping (V-SLAM), inverse rendering, and other related computer vision techniques known by different names all observe 2-dimensional imagery and estimate 3D structure, as shown in Figure 2. The current state of the art V-SLAM techniques can run on live video in real-time, but the 3D mapping quality is sparse in detail. When high-quality mapping is instead the focus, modern photogrammetry techniques must instead run on a static corpus of data. Even when run on modern computing infrastructure, these can take hours to estimate a high-quality 3D map, and even longer for larger areas. There exists a technology gap for real-time 3D mapping on live video, especially when considering multiple-agent perspectives. Overcoming these current limitations would enable new applications for MR and military training applications.

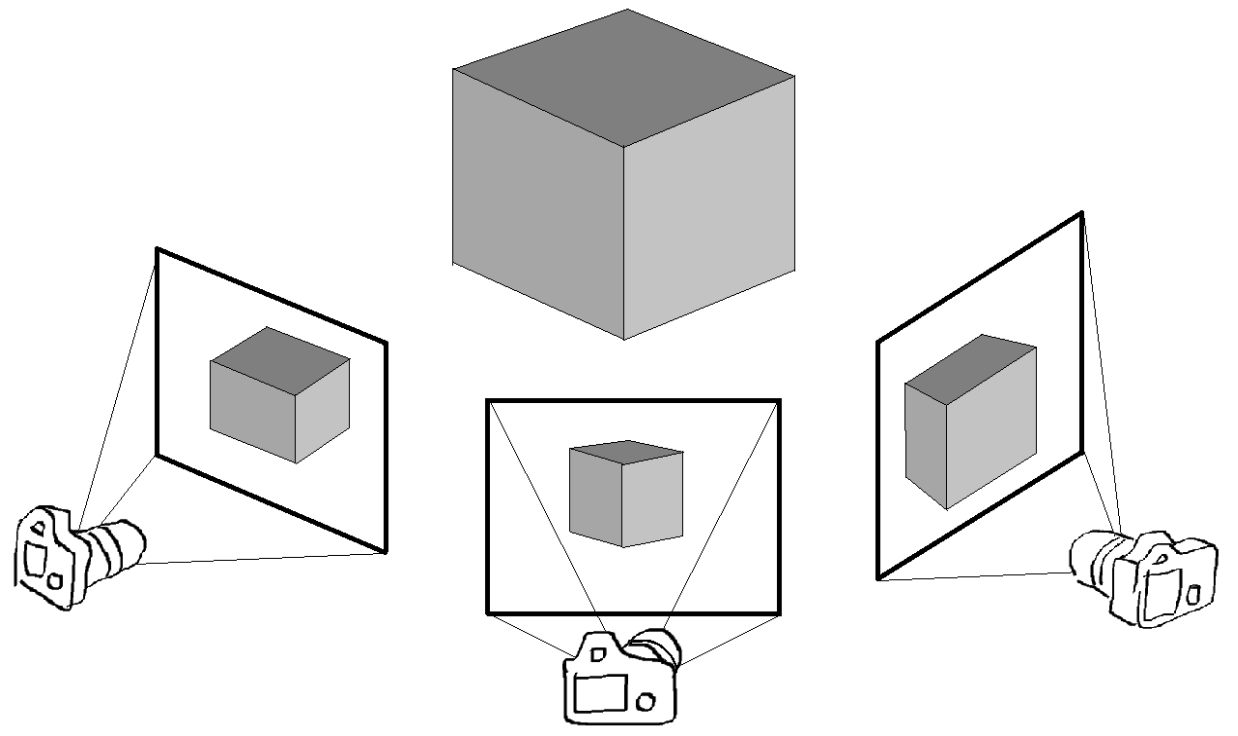

**Figure 2. Multi-view geometry calculates 3D structure from 2D images from different perspectives.**

### Structure from Motion (SfM)

Photogrammetry often shares its technical foundations with the academic nomenclature structure from motion and multi-view geometry (Hartley & Zisserman, 2004). The SfM problem has also been studied for many years , with powerful partial solutions available in the last decade (Ozyesil et al., 2017). Described in the most general case: Given a collection of images, calculate the 3D structure contained in the scene. The images from single camera photographs or video must include some motion through the 3D space to gain multiple viewing angles of the environment (Braun et al., 2013), often with the most success possible when camera travel completes a loop around an object. This technique can produce dense 3D model estimates, but the process typically has a long run-time and usually only interacts with a static set of data. It can be challenging to register new image information to an existing 3D solution. These limitations prevent SfM from running in real-time. Instead, SfM is often focused on 3D reconstruction visual quality. Often photogrammetry is concerned with an individual object as the focus, but SfM can also reconstruct less-structured environments. Even with photogrammetry finding great success in common practice, there are surprisingly few datasets and evaluation metrics for large-scale outdoor applications (Bianco et al., 2018). This is because scenes of the real world must be measured with lidar to capture an imperfect 3D ground truth. Dynamic objects and ever-present environmental effects from outdoor conditions make ground truthing a 3D scene challenging. The most widely recognized SfM algorithm of the last decade is explored.

**COLMAP**

The landmark "Structure-from-Motion Revisited" paper (Schonberger & Frahm, 2016) and accompanying open-source application set a high standard for traditional SfM and multi-view geometry pipelines. COLMAP, abbreviated from "collection mapper", is the baseline standard for offline SfM techniques. Although the technique offers an incremental reconstruction process for video imagery, it is best known for a pre-computed sort of unconstrained collections of photographs. For this use case, no prior information is known of the photographs, and an initial search must be performed to select a pair of visually overlapping initialization images. All images must perform visual feature extraction, a feature correspondence search, and geometric verification to select two anchor images with sufficient overlapping scenes. For our real-time application focused on video, some of these initialization steps are not as applicable, since image overlap should exist from one video frame to the next. COLMAP begins with feature extraction using the Scale Invariant Feature Transform (SIFT). This identifies patterns in the image which can be matched between images. Next, the technique builds out a point cloud map. This includes overlapping-image registration, matching and adding to points in the cloud, triangulation of each new image's camera position, and bundle adjustment to refine both points and camera poses. These steps are repeated for each additional image, either selected for similarity of features in the cloud, or time sequential video frames in our case. This baseline technique had a reported run-time of approximately 2 seconds per frame (Schonberger & Frahm, 2016), but we measured this to be significantly longer in practice on our evaluation system. Significant research has preceded and followed this seminal paper, including techniques for robust and computationally efficient features, search strategies, triangulation, and recursive optimization. Below we demonstrate an outcome of the open source COLMAP software package. In Figure 3 below, Sequence 4 of the KITTI Odometry dataset was processed to produce a color point cloud and the accompanying camera positions in red.

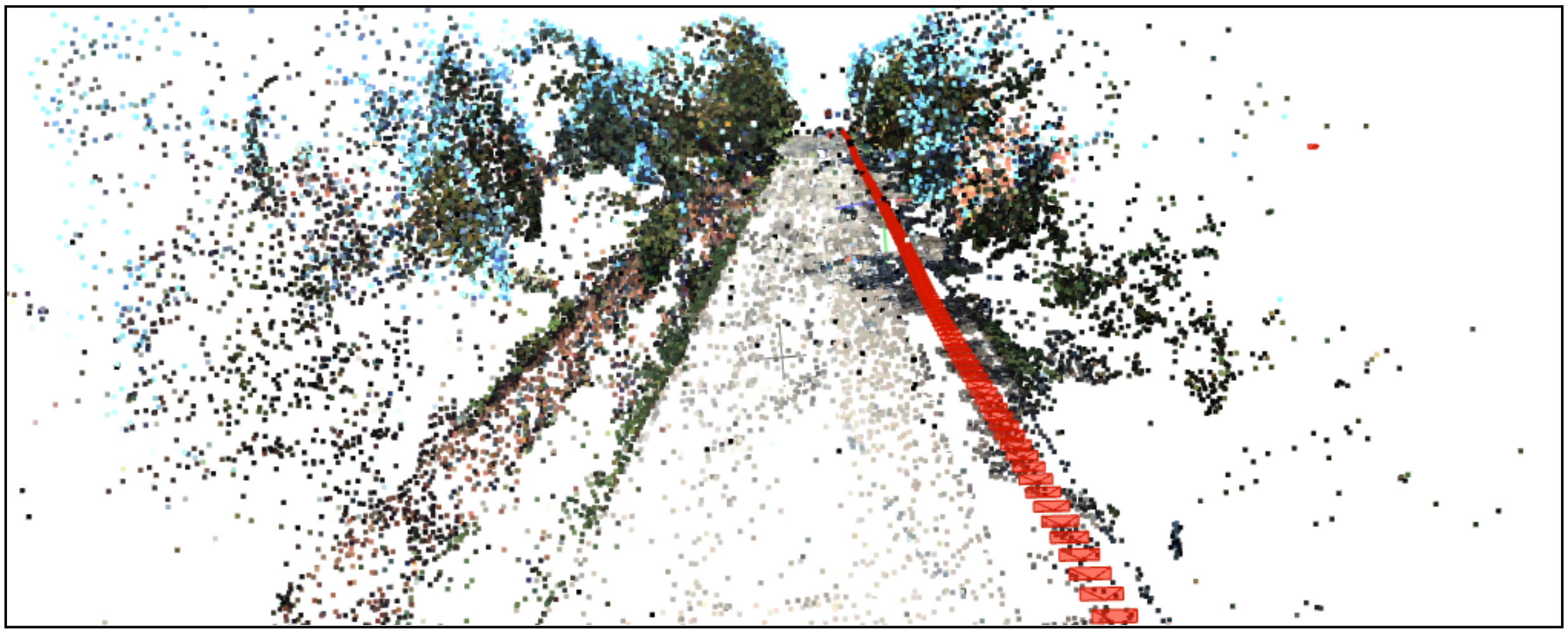

**Figure 3. COLMAP algorithm visualization of the KITTI scene. Generated from open source code.**

**Simultaneous Localization and Mapping (SLAM)**

SLAM has been studied for many years related to robotics (Aulinas at al., 2008). An autonomous robotic agent must both map the local area around it and track its position within that map. SLAM often tracks point cloud data from a lidar sensor measuring range in 3D. For these robotic applications, a dense point cloud map of an environment is not required: if a sparse set of points can map and track nearby objects, this is enough to alert the robotic agent of an obstacle. Speed is the highest priority for most SLAM systems (Mur-Artal et al., 2017), with real-time calculations required for most robotics applications. Few SLAM systems consider real time measurements from multiple agent perspectives (Chang et al., 2020), which is required for a large-scale digital twin model of the world.

In contrast, visual-SLAM (V-SLAM) uses only 2D video imagery to calculate 3D point-clouds (Fuentes-Pacheco et al., 2015). V-SLAM and SfM follow similar processing steps: simultaneously estimate the positions of the camera locations while calculating the 3D structure of an environment map; repeat this process with more imagery; and iteratively refine all estimates. These calculated 3D point clouds are created sequentially from streaming video, but the quality of this 3D data is lacking the dense detail needed for other mapping applications (Engel et al., 2014; Engel et al., 2015). V-SLAM is engineered for robotic navigation, but for human in the loop applications, like AR, higher fidelity is needed (Huang et al., 2020). Higher fidelity is often realized by leveraging more sensors in tandem through sensor fusion. A robotic application with cameras and lidar together is often called the “RGB-Depth” (red, green, blue colors) case of V-SLAM. Very commonly V-SLAM also leverages inertial measurements from sensors, referred to as “visual inertial SLAM” (VI-SLAM) (Piao & Kim, 2017). Whenever multiple cameras are available on the same rigid platform with consistently overlapping perspectives, the term “stereo vision” is used. Finally, when only position and navigation measurements are required without a persistent map, this solution simplifies to “visual inertial odometry” (VIO). This is most common for autonomous or semi-autonomous vehicles and for GPS-denied navigation. Very few open source packages attempt to do V-SLAM, VI-SLAM, RGB-D, and RGB-D with inertial measurements, and stereo vision versions of all these varieties, except one.

**ORB-SLAM3**

ORB-SLAM3 (Campos et al., 2021) was recently published offering a combined solution for many potential data types as inputs, and it is a best-in-class solution for quickly tracking large scenes across multiple maps. This is the most recent work by the University of Zaragoza (Mur-Artal & Tardos, 2017; Elvira et al. 2019). ORB-SLAM3 is one of the fastest and the most feature-rich V-SLAM open-source libraries. It includes a large variety of supported data input types, including single camera, stereo cameras, omnidirectional cameras, depth sensing lidar, accelerometers, and inertial measurement unit (IMU) hardware sensors. The ORB-SLAM3 algorithm as published calculates and tracks new video frames in 22 milliseconds on average (Campos et al., 2021). This computational time increases with the data types used as input and the quantity of 3D feature points in a map. The most computationally expensive operation is the global point cloud mapping and optimization process, which requires 267 milliseconds on average to complete each step but can easily double in time for large scenes (Campos et al., 2021). ORB-SLAM3 is one of the few SLAM techniques to support an "atlas" of historical maps which can be fused to new measurements (Elvira et al. 2019). In our evaluation, only one camera was used without the additional supplementary data inputs. We demonstrate an outcome of the open source ORB-SLAM3 software package. In Figure 4, the driving Sequence 4 of the KITTI Odometry dataset was processed to produce the estimated camera positions in blue and the red cloud of 3D map points.

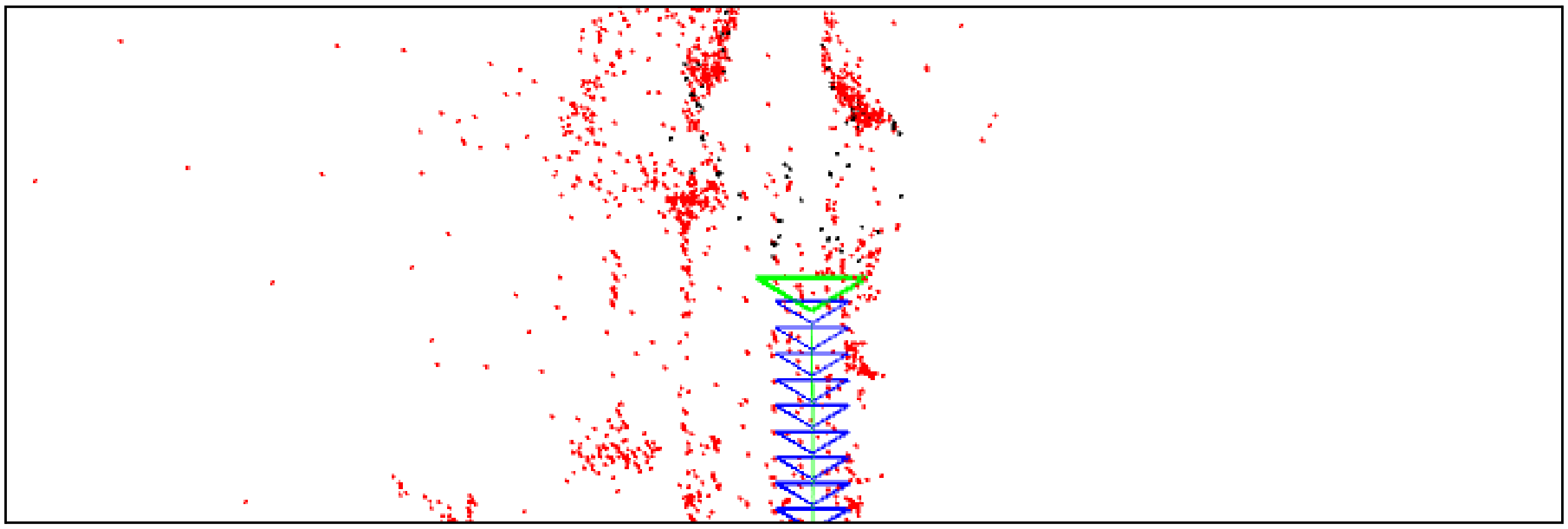

**Figure 4. ORB-SLAM3 algorithm visualization of the KITTI scene. Generated from open source code.**

**ORB Features**

ORB-SLAM relies upon the popular “Oriented FAST and Rotated BRIEF” (ORB) feature points (Rublee et al., 2011). ORB modifies the “Features from Accelerated Segment Test” (FAST) edge-detection technique and the “Binary Robust Independent Elementary Features” (BRIEF) descriptors. These techniques were created as an open-source alternative to other patented feature point types. ORB features have been evaluated to be computationally efficient by trading matching accuracy for resistance to sensor noise. Each ORB point on an image has a location, orientation, scale, detected edge-strength, and the modified-BRIEF binary vector identifier. Typically,

ORB features are calculated for grayscale imagery in order to optimize for speed, but ORB features can also run on video images with the red, green, and blue (RGB) color space. This allows the point to include a color value when plotted. In Figure 5 of the same sequence as above, the corresponding tracked feature points are shown in green.

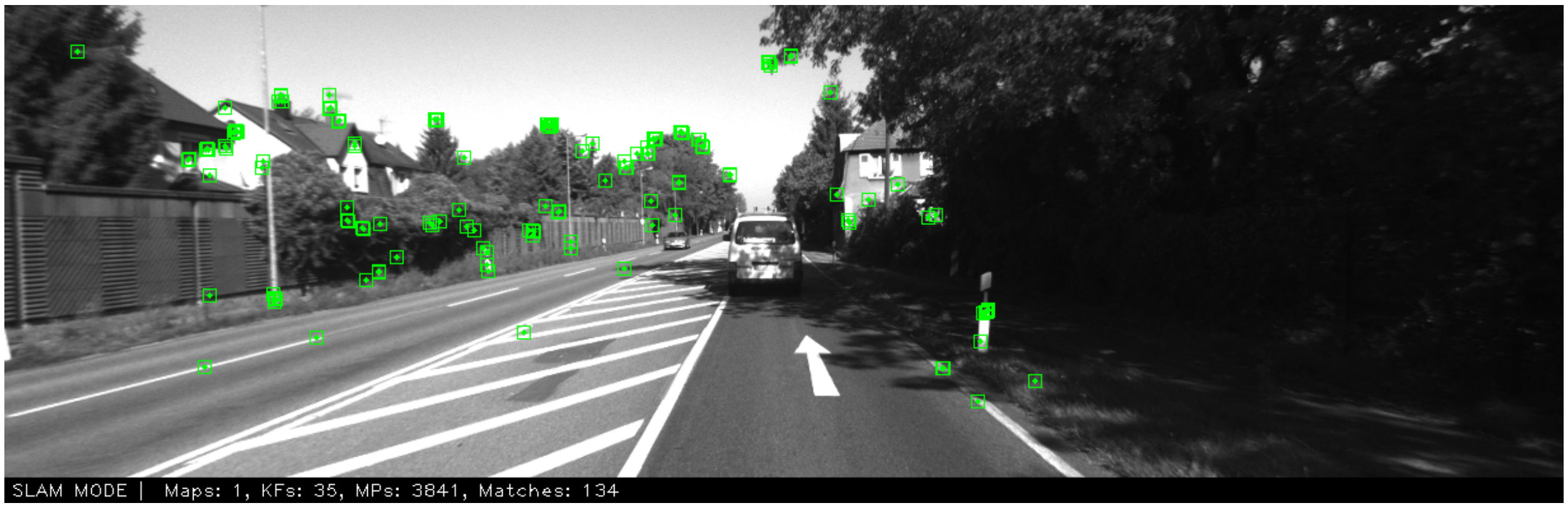

**Figure 5. ORB feature points track unique patterns in KITTI video. Generated from open source code.**

**Neural Radiance Fields (NeRF)**

The limitations of V-SLAM and SfM are an active area of research for Deep Neural Network (DNN) based techniques (Chen et al., 2020). Techniques such as NeRF (Mildenhall et al., 2020; Martin-Brualla et al., 2021) have set a new standard for SfM benchmarks, but these techniques still suffer from the same limitations as traditional SfM solutions: all data must exist before the reconstruction algorithm can begin, the algorithm relies upon known or estimated camera positions, and the solution takes some time to optimize. This is discussed in greater detail in the following sections. The latest techniques released in the last year show how new DNN learning methods, often called Neural Rendering, can be run in real-time to generate dense 3D visuals. As these performance issues are solved, a vast new set of capabilities could be enabled for many visual graphics applications.

The many related NeRF techniques are a dramatically new way to create 3D data from image collections, very similar to photogrammetry. Authored by Berkley and Google researchers, these NeRF ray tracing techniques render novel views of a scene and, with additional processing, for estimating scene 3D geometry. A NeRF method includes the following steps:

1. Begin with a ground truth source image from a 3D location point of view.
2. The algorithm traces geometric rays through each pixel of the image into 3D space.
3. Upon each ray, it considers occlusion and sums color (radiance) learned by the DNN.
4. This process estimates the aggregate color information for each pixel.
5. Many trace rays, for many pixels, together form a new rendered image estimate.
6. The DNN learns to minimize error between the ground truth and the estimated image.
7. Repeat for all images used as inputs for training.

NeRF learns to represent the color radiated from different 3D locations in space. This work is extended by "NeRF in the Wild: Neural Radiance Fields for Unconstrained Photo Collections" (Martin-Brualla, 2021) and work by many other fascinated researchers. These improved techniques can render fully-realized 3D visual scenes, but this comes with several drawbacks. Most relevant for mapping, NeRF architectures can only calculate 3D structure through the lines which span a grid through 3D space. This is expensive to compute a full 3D geometry mesh, and mostly incompatible with traditional 3D graphics rendering strategies for textures, polygons, point clouds, and voxels. Although these techniques have only been explored for a static set of images which can be fully learned by a DNN, it may be possible to retrain these techniques incrementally with new imagery inputs, thereby updating structure with a newly-observed environment state. In a recently published work, the Block-NeRF technique (Tancik et al., 2022) was used to reconstruct a city at neighborhood scale. Neural rendering techniques are showing increasing performance at scale, while reducing speed and approaching real-time performance.

**Instant Neural Graphics Primitives**

Recently NeRF has experienced a dramatic increase in interest due to publicity from Nvidia (Müller et al., 2022). Nvidia released a software implementation for training and rendering of NeRF called "Instant Neural Graphics Primitives" (Instant-NGP). This makes extensive use of Nvidia graphics processing units (GPU) and the compute unified device architecture (CUDA) accelerated processing instruction set. Nvidia markets the Instant-NGP open source code as "training in minutes, rendered in seconds". We evaluated these speed claims for 3D reconstruction. Given known or pre-computed camera positions, we observed Instant-NGP able to train a NeRF in seconds, with a high-quality rendering taking longer. Due to the incremental DNN training procedure of this neural rendering technique, the 3D reconstruction quality increased with additional compute time. Figure 6 demonstrates the visual quality under our limited real-time video conditions. The speed of this technique opens new possibilities for 3D reconstruction and digital twins.

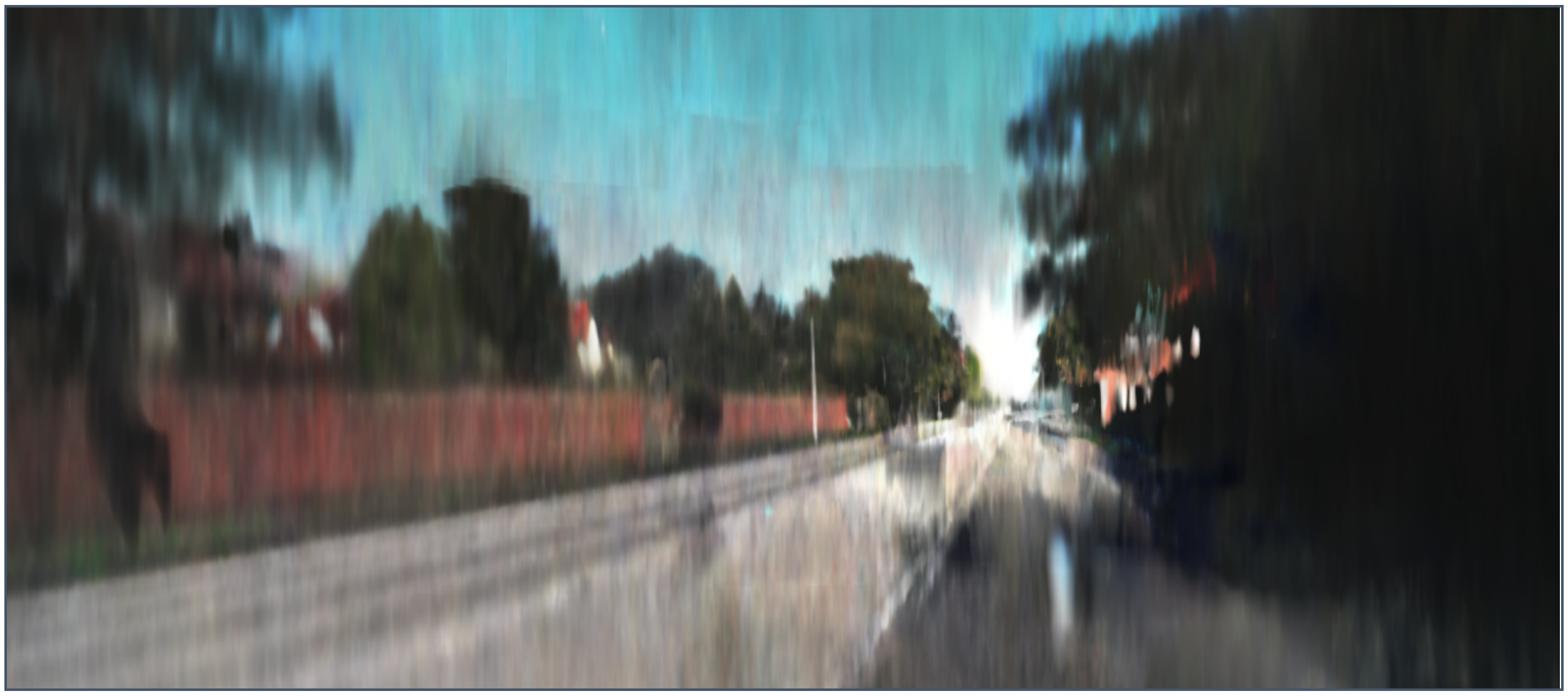

**Figure 6. NeRF algorithm visualization of the KITTI scene. Generated from open source code.**

**EVALUATION**

We evaluated the three algorithms, COLMAP, ORB-SLAM3, and NeRF, on test conditions similar to military training scenarios. For this study we focused on outdoor and large-scale environments. Besides video from ground-level perspective, we also needed to know the approximate ground truth of the 3D environment. Lidar is perhaps the only way to procure this 3D information for comparison against each videogrammetry technique. Our evaluation was focused on applicability for real-time 3D reconstruction for digital twins, both speed and quality. Since the algorithms produce substantially different outputs, it was challenging to evaluate 3D quality quantitatively. Nonetheless we intend to publish these findings in future publications.

Due to the scarcity of outdoor datasets with lidar measurement information, we utilized the KITTI Vision Benchmark Suite vehicle driving dataset (Geiger et al., 2013). We specifically used the KITTI Visual Odometry / SLAM Evaluation data published in 2012. This includes video camera imagery, 3D point cloud information, GPS and IMU conditioned camera positions, with everything synchronized at 10 recordings per second (Hz). There are 22 recorded driving sequences within this dataset; often these are split between 11 training and 11 test sequences. Each sequence has various driving time, distance, speed, and terrain. Using this dataset and the previously discussed algorithms, we demonstrated the qualitative results, with quantitative results remaining for future publication.

An example of the KITTI dataset is shown in Figure 7. A video image from KITTI Sequence 4 is displayed with lidar distance information overlaid. The colorful points show lidar depth measurements as hot colors at near distance and cold colors when further away. Only the lower hemisphere of the image includes corresponding lidar data. The lidar sensor is limited by both low angle directions and overall range distance. The camera has a particularly wide field of view in the horizontal direction.

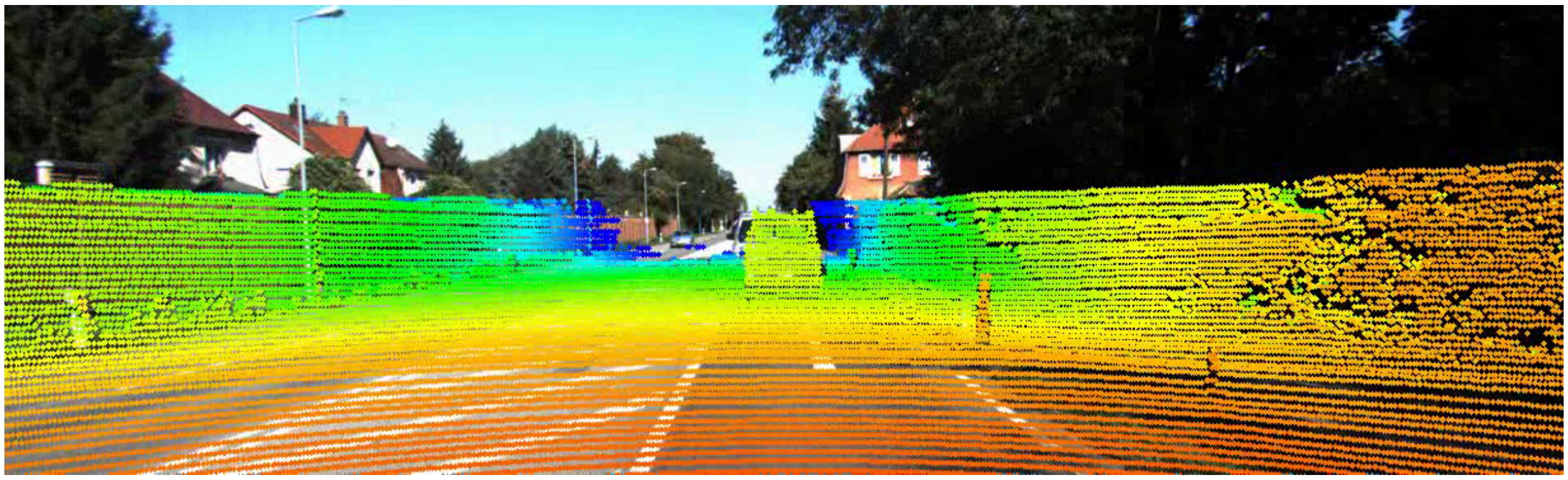

**Figure 7. Lidar measured ground truth distances on the KITTI scene.**

Our evaluation was done on the same computing system consisting of: an Intel Core i7 9th generation CPU with 6 physical processing cores, 16GB of random access memory, an Nvidia GeForce RTX 2060 mobile GPU with 6GB of memory, the latest generation solid state drive, and Ubuntu 20.04 LTS. These hardware specifications describe a mid-range gaming-capable mobile computer produced in 2019. These computer specs may be comparable to the latest generation of Nvidia Jetson products or other AI/ML purpose built small form-factor devices.

## RESULTS AND DISCUSSION

With the evaluation methods presented, we tested the COLMAP, ORB-SLAM3, and NeRF techniques against the KITTI video camera data. Table 1 groups algorithm steps and computational time per frame of video. In the sections that follow, we detail the qualitative results of the algorithms.

**Table 1. Computational Time**

| | **Processing Time Per Video Frame (in seconds)** | | | | | |
|---|---|---|---|---|---|---|
| | **Total (Ideal)** | **Feature Creation** | **Feature Matching** | **Camera Pose** | **3D Model** | **Live 3D Render** |
| **COLMAP (CPU Only)** | **1.33** | 0.41 | 0.061 | 0.86 [1] | | N/A |
| **ORB-SLAM3 (CPU Only)** | **0.024** | 0.024 | | | | 0.00 [2] |
| **Instant-NGP NeRF (GPU)** | **0.590 [3]** | N/A | N/A | 0.080 [4] | 0.40 | 0.11 |

[1] *Ideal conditions, not including optimization non-convergence.* [2] *Includes live 3D render at negligible time impact.*
[3] *Measured in seconds per training/render, not frames.* [4] *Optimization of pose, but relies on existing estimates.*

### COLMAP

The thorough SfM algorithm includes several steps in its complete process. This includes SIFT feature extraction, feature matching between images, camera localization, and 3D reconstruction. The first distinct step is to calculate and create SIFT feature points. This utilized 12 CPU threads maxing out the system. The application warns against using high resolution imagery due to memory requirements, but we observed only a 1.1GB peak memory usage at this stage when utilizing the KITTI video frames of 1226 by 370 pixels. Next, each feature descriptor is matched compared to all other feature points. Due to the efficient search strategy, this process completes quickly. Finally, COLMAP registers each new frame of video estimating the new camera's position compared to the 3D cloud of feature points. This optimization problem adjusts cameras and 3D points simultaneously making each step hard to quantify individually. This optimization CPU operation ran primarily in one processor thread. Infrequently there were several processes running in parallel, although the entire CPU was never fully utilized. Often COLMAP would encounter faults or require recalibration. The camera pose / 3D model metric above is under ideal conditions not including these slowdowns, per note [1]. In practice COLMAP takes much longer than the total ideal processing time per video frame listed. Additionally, it does not always come to a successful 3D model, with initialization errors being a primary culprit. With additional engineering and configuration it may become faster than the listed performance, especially if the software can be compiled to utilize the GPU.

**ORB-SLAM3**

This SLAM technique is known for being fast. Without breaking apart the overall time per frame into subcategories, it clearly outperformed the other two 3D reconstruction algorithms in terms of speed. This implementation was compiled without CUDA, and runs solely on only one CPU core. ORB-SLAM3 tracks only a minimal number of 3D points in the digital world model, as can be seen in Figure 4. In practice, this technique alone is insufficient for generating a large-scale 3D digital twin. It may be possible to combine the speed of this technique with the dense 3D visual quality of other techniques.

**Instant-NGP NeRF**

Neural rendering has improved quickly in the last few years, with speed approaching real-time for edge computing when a GPU is available. Currently this technique does not execute in time sequential order. Instead it takes all video images at once, incrementally training the neural rendering technique with several images at once per batch. We believe this paradigm can be modified to allow for new, live video imagery. The overall speed metric for Instant-NGP NeRF is different from the other two algorithms. The speed of NeRF is dependent upon neural network training and rendering, not the rate of processing the next frame, per note [3]. Note that NeRF still relies upon camera pose locations from another source. The developers of Instant-NGP bundle COLMAP to calculate camera positions, but this could instead be captured for hardware sensors or a visual odometry algorithm like ORB-SLAM. Overall, we found the visual quality of NeRF to be not yet perfect, but a substantial improvement over traditional geometric 3D reconstruction. The quality of NeRF may be improved with more video image frames, photographs with less motion blur, and a predetermined scale for world size. Finally, NeRF requires additional processing to extract a 3D model for a digital twin. We have not yet benchmarked the processing cost to extract 3D polygons.

## FUTURE WORK

KITTI is a valuable dataset for evaluation of which we have only scratched the surface. In the future, we intend to evaluate the placement of 3D points as compared to the ground truth lidar measurements. The 3D reconstruction on the whole can be compared to the lidar measurements across a driving sequence. Recently an equally substantial dataset known as KITTI-360 was published by Liao et al. (2021). This included 360 degree views around the vehicle, with the camera imagery also aligned to lidar measurements. This is another valuable resource for measuring 3D point placement accuracy.

Besides utilizing popular academic datasets for evaluation, we chose to stress test these algorithms in especially challenging scenarios. In the future, we will publish an evaluation of these algorithms on video collected during military simulation paintball events. This test case demonstrates the algorithm in a fast-paced combat environment. This data source is at limited standoff-distances though, and not at the long distances common to military firefights. The fast movements of the helmet mounted camera can cause ORB-SLAM3 to lose the track of the 3D features. COLMAP and NeRF to not consistently calculate a successful solution on this stress test. In another planned benchmark, we have collected imagery from Grand Canyon national park from body camera video. These videos include a limited set of popular hiking trails. This test requires more engineering effort before publishing.

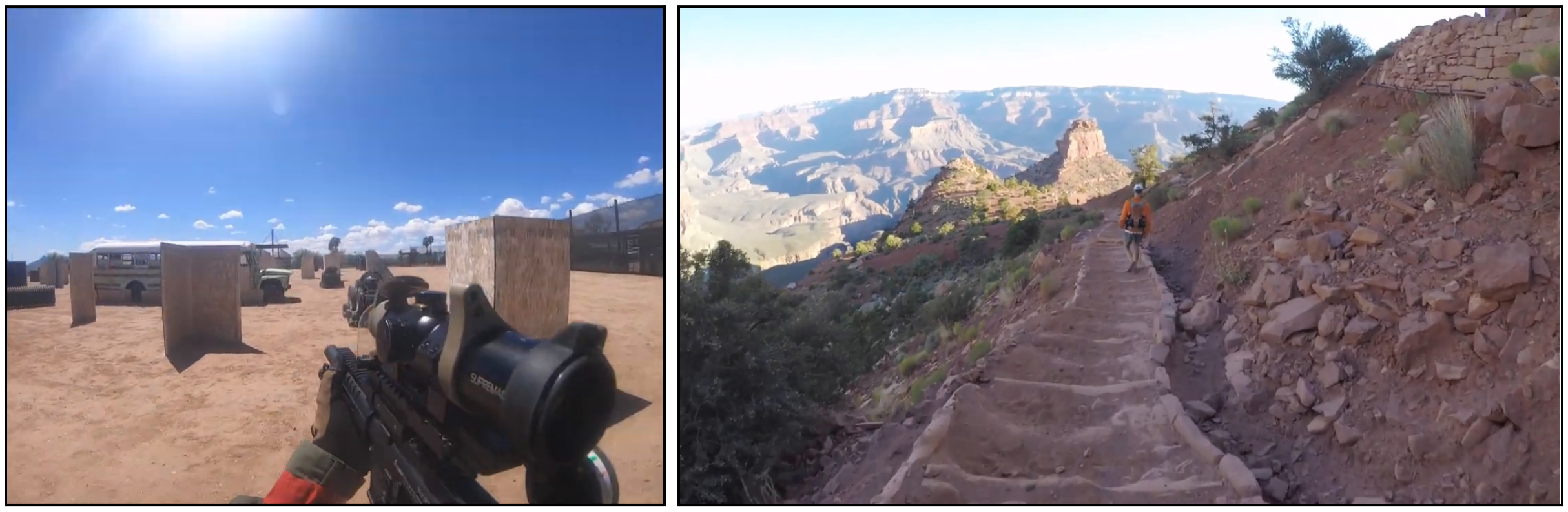

**Figure 8. *Military-Simulation Paintball* and *Grand Canyon Hike* future benchmarks.**

## CONCLUSION

Mass adoption of MR seems near for both civilian and military applications. Smartphones have enabled AR to visualize objects, but rarely do visualization include a full understanding of the environment. We have demonstrated the utility of 3D reconstruction algorithms to produce a digital twin in real-time. We believe MR in the training field can be enabled through the use of the IVAS system, 3D reconstruction algorithms, 5G wireless networks, and a defense industrial base willing to embrace consumer technologies. Doing so will control the development costs of such systems while enabling an enhanced method for soldiers to train against a realistic enemy.

## ACKNOWLEDGEMENTS

We acknowledge General Dynamics Mission Systems for supporting in part this publication. Collaboration with and investment by the Training Test and Efficiency Solutions (T2ES) group has been pivotal to the understanding of this problem space. This project was funded in part by the NSF CISE awards 1659871 and 1953745. We appreciate the research assistance of Ebubekir Şen and Emilio Montoya, through the SenSIP Center research experiences for teachers and research experiences for undergraduates programs. We appreciate the open source collaboration from Dominique Warner.